\documentclass[twocolumn,secnumarabic, graphics,floatfix,nofootinbib,nobibnotes,aps,prb,10pt]{revtex4-1}
\usepackage[english]{babel}
\usepackage{epsf}
\usepackage{epsfig}
\usepackage{graphicx}
\usepackage{xcolor}
\usepackage{ulem}
\DeclareUnicodeCharacter{2212}{\ensuremath{-}}
\begin{document}

\title{$Ab-initio$ Prediction of Mechanical, Electronic, Magnetic and Transport Properties of Bulk and Heterostructure of a Novel Fe-Cr based Full Heusler Chalcogenide} 

\author{Joydipto Bhattacharya$^{1,2}$, Rajeev Dutt$^{1,2}$, Aparna Chakrabarti$^{1,2}$}

\affiliation{$^{1}$ Raja Ramanna Centre for Advanced Technology, Indore-452013, India}

\affiliation{$^{2}$Homi Bhabha National Institute, Training School Complex, Anushakti Nagar, Mumbai-400094, India}

\begin{abstract}
	Using electronic structure calculations based on density functional theory, we predict and study the structural, mechanical, electronic, magnetic and transport properties of a new full Heusler chalcogenide, namely, Fe$_2$CrTe, both in bulk and heterostructure form. The system shows a ferromagnetic and half-metallic(HM) like behavior, with a very high (about 95\%) spin polarization at the Fermi level, in its cubic phase. Interestingly, under tetragonal distortion, a clear minimum (with almost the same energy as the cubic phase) has also been found, at a c/a value of $\sim$1.26, which, however, shows a ferrimagnetic and fully metallic nature. The compound has been found to be dynamically stable in both the phases against the lattice vibration. The elastic properties indicate that the compound is mechanically stable in both the phases, following the stability criteria of the cubic and tetragonal phases. The elastic parameters unveil the mechanically anisotropic and ductile nature of the alloy system. Due to the HM-like behavior of the cubic phase and keeping in mind the practical aspects, we probe the effect of strain as well as substrate on various physical properties of this alloy. Transmission profile of the Fe$_2$CrTe/MgO/Fe$_2$CrTe heterojunction has been calculated to probe it as a magnetic tunneling junction (MTJ) material in both the cubic and tetragonal phases. Considerably large tunneling magnetoresistance ratio (TMR) of $\approx 10^3$ \% is observed for the tetragonal phase, which is found to be one order of magnitude larger than that of the cubic phase.
\end{abstract}
\maketitle

\section{Introduction}
Half-metallic (HM) ferromagnets (FM) have become a topic of active research due to their potential for various technological applications. Theoretically, the HMFM materials are shown to exhibit 100\% spin polarization (SP) at the Fermi level (E$_{F}$), with one of the spin channels showing semi-conducting (SC) and the other one possessing metallic nature. In 1983, half-metallicity has been predicted in some half Heusler alloys (HHA), namely, NiMnSb and its isoelectronic compounds, PtMnSb and PdMnSb.\cite{Groot_1983} Ever since this study has been published in the literature, the field of HM Heusler alloys (HMHA) has attracted immense attention of the researchers, both theoreticians and experimentalists alike. As a result, innumerable studies on half-metallic half as well as full Heusler alloys (FHA) have come to the fore, in order to either understand some fundamental aspects or explore their potential for various applications.\cite{WEBSTER_1971,Miura_2008,Peter_2009,Roy_2016,ENAMULLAH20181055,Baral_19,KANG20179,KERVAN20111358,ZHU2014391} 

It has been seen in the literature that typically many of the FHAs exhibit metallic nature, while a large number of HHAs are SC in nature. Many works have been carried out to show that the number of valence electrons (n$_v$) plays a crucial role in defining the electronic as well as magnetic properties of materials. The Slater-Pauling rule established the relation between n$_v$ and the magnetic moment of a transition metal.\cite{Slater_1936,Pauling_1938} Further, for Heusler alloys, specifically for Co-based ones, Slater-Pauling behavior has been seen, as reported in the literature.\cite{Galanakis,Luo_2008}  It has been observed that among the FHAs, primarily Co-based alloys show HM behavior with a typical value of valence electrons (n$_v$) of 26 to 28. As discussed above, searching for a new or novel HMHA, which has high to very high (preferably 100 \%) spin polarization at E$_F$, recently become of utmost importance, both from the points of view of fundamental understanding and technological application.\cite{Hirohata2013,Kelvin,Bainsla} To this end, we take a combination of Fe, Cr and Te atoms, yielding a FHA, Fe$_2$CrTe, with a n$_v$ value of 28. This alloy contains no Co atom, on the other hand, contains a chalcogen atom,Te. In the recent past, chalcogen atoms have been shown to be elements of interest\cite{DUTTRAJEEV,PANDEY_2019,PANDEY_2021,Ru_2006,Zhu_2016}. From our $ab-initio$ electronic structure calculations, it turns out that Fe$_2$CrTe alloy is expected to show a HM-like behavior. The HM properties have been observed to be greatly influenced by defects, surfaces and interfaces. In this context, the study of the effect of strain as well as substrate on various physical properties of a HMHA can be interesting and important from practical application points of view. Hence, we embark upon the same in the present work. We apply uniform isotropic strain and also apply a bi-axial strain by putting the cubic alloy on a well-known and lattice-matched SC substrate, namely, MgO.

In one of our recent works, we have probed the possibility of coexistence of half-metallicity and tetragonal (martensite) transition in a series of Ni- and Co-based FM FHAs, including Ni$_2$MnGa and a few other well-known alloys.\cite{Roy_2016} As a martensite phase transition (MPT) indicates occurrence of a shape memory behavior in a magnetic alloy and a HMFM alloy has the possibility of application in the field of spintronics, studying both these aspects is important. Since typically it has been observed that the MPT and HM behaviors are not seen in the same material, we have dwelved in the study of the same in the past.\cite{Roy_2016} We predicted a novel Co-based FHA (Co$_2$MoGa), which exhibited a tendency of MPT and also a HM-like behavior. Three of the other studied FHAs have shown a very clear local minimum in the energy versus c/a plot, with a range of values of c/a ratio ($\sim$1.25 to 1.35). While the clear display of a tetragonal (martensite) phase is well-known and it is well-studied in many of the FHAs\cite{Roy_2016,Chakrabarti_2013,Barman_2005,DUCHER2008213,Siewert_2011,ROY2016929}, the appearance of a minimum at a c/a ratio other than 1, in an otherwise cubic alloy, has seldom been observed.\cite{Roy_2016} In this work, we explore the possibility of a tetragonal distortion for the FHA Fe$_2$CrTe and find that along with a cubic (austenite) phase, a clear minimum is observed for a tetragonal phase with a c/a ratio of $\sim$ 1.26, leading to a double-minima like structure in the energy versus c/a plot.

For spintronic devices, in recent times, an extensive search for new materials suitable for magnetic tunneling junction (MTJ) is going on.\cite{Feng_2022,D1CP01579F} These heterojunctions comprise two FM electrode materials and a non-magnetic insulator or semi-conducting spacer material in between the two electrodes. In these systems, the tunnenling magnetoresistance (TMR) is strongly dependent on the relative spin orientations of both the electrodes (parallel or anti-parallel). The TMR ratio has been defined as the difference in conductance of the MTJ in two different magnetic orientations divided by the smaller value. This ratio can in principle (theoretically) be infinitely large if a HMFM material is used as an electrode in the MTJ. In the literature, many studies have been reported where Co-based HMFM alloys have been used. Most of these studies include a thin insulating layer of MgO as the barrier/spacer material and favorable tunneling properties have been observed in these MTJs.\cite{Miura_2011,Peter_2009,HAN201795,Gercsi,ROY2020166092} As Fe$_2$CrTe is expected to exhibit a HM-like character, we first probe the magnetic properties of the Fe$_2$CrTe thin film (13 monolayers (ML)) with 5 (7) ML of MgO as a substrate material. We find that the SP at E$_F$ is about 75\% (70\%), when 5 (7) ML of MgO substrate. We further calculate the transmission properties of the heterojunctions Fe$_2$CrTe/MgO/Fe$_2$CrTe to understand and explore the potential for MTJ application.

In the next section, we discuss the method of electronic structure calculations, which is based on density functional theory. We also briefly discuss the calculational method related to transport properties. In the section followed by methodology, we present our results and discuss the same. Finally, we summarize and conclude our work in the last section.

\section{Method}  
The FHAs are known to exist either in a conventional or an inverse Heusler alloy structure. From the structure optimization, we find that in the lowest energy state, Fe$_2$CrTe alloy possesses the conventional structure with a $L2_{1}$ phase that consists of four interpenetrating face-centered-cubic (fcc) sub-lattices with origin at the following fractional positions: Fe atom at (0.25,\,0.25,\,0.25) and (0.75,\,0.75,\,0.75) sites, Cr atom at (0.5,0.5,0.5) site and Te atom at (0,0,0) site. The structure has been optimized by doing full geometry optimization using Vienna Ab Initio Simulation Package (VASP)\cite{Kresse_1996,Kresse_1999} with the projector augmented wave (PAW) method.\cite{Kresse_1996,Kresse_1999}  For exchange-correlation (XC) functional, generalized gradient approximation (GGA) over the local density approximation has been used.\cite{Perdew} We use an energy cutoff of 500 eV for the planewaves. The final energies have been calculated with a $k$ mesh of 15$\times$15$\times$15 for the cubic symmetry and an equivalent number of k-points for the tetragonal symmetry. The energy and force tolerance for our calculations were 1 $\mu$eV and 20 meV/\AA, respectively. For obtaining the electronic properties, the Brillouin zone integration has been carried out using the tetrahedron method with Bl\"ochl corrections. The directional dependencies of different mechanical properties (Young's modulus, inverse of bulk modulus or compressibility, shear modulus, and Poisson's ratio) of this alloy in both the cubic and tetragonal phases have been calculated with the help of the ELATE software.\cite{MARMIER20102102} 
For the calculation of transport properties of the heterojunction of Fe$_2$CrTe and MgO, we make use of the PWCOND code\cite{Smogunov_2004}, which has been implemented in the Quantum ESPRESSO (QE) package\cite{Baroni_2001}. The spin-dependent tunneling conductance has been calculated using the Landauer formula:\\
\begin{center}
        $G^{\sigma} = \frac{e^2}{h}\sum_{K_{||}}T^{\sigma}(K_{||},E)$
\end{center}
here $\sigma (= \uparrow,\downarrow$) is the spin index and $T^{\sigma}(K_{||},E)$ is the spin-dependent transmission coefficient at a particular energy value $E$, with $K_{||} = (K_x, K_y)$, where $K$ are the wave-vectors, x and y are directions.

In the literature, Choi and Ihm\cite{Choi_1999} have given a method to calculate $T^{\sigma}(K_{||},E)$. We perfom this calculation for the optimized geometry using the GGA exchange functional.\cite{Perdew} We have taken the cut-off energy for the wave function and the charge density as 60 and 600 Ry, respectively. A mesh of 12$\times$12$\times$1 k-points have been used for the self-consistent-field(SCF) calculation of Fe$_2$CrTe/MgO/Fe$_2$CrTe heterojunction. A high tolerance (10$^{-8}$ Ry) and a large k-mesh (100 points in both $x$ and $y$ directions) have been taken, which are required to capture the fine spikes in a transmission.\cite{Karki_2021}
The scalar relativistic ultrasoft pseudopotentials(USPP) with the GGA exchange-correlation term have been used, as obtained from the PSLibrary 1.0.0. For further details of calculation of ballistic conductance see Ref.\cite{Choi_1999}. Convergence of all the relevant parameters for VASP and QE packages have been tested before embarking upon the calculations of the physical properties.

\section{Results and Discussion}
In this section, first we predict the energetic stability of the cubic phase of Fe$_2$CrTe alloy and then explore the possibility of a stable tetragonal (martensite) phase of this alloy, where the c/a ratio (a and c being the lattice constants along the x and z-directions) has been varied, keeping the volume fixed. We then calculate and present the dynamical and mechanical properties of these two phases. The mechanical properties of the well-known HA Ni$_2$MnGa have been discussed in some places, for the sake of comparison. Further, we discuss the electronic and magnetic properties of the cubic and tetragonal phases. Due to the near HM behavior of the cubic phase, keeping in mind the practical aspects, we probe this phase further. An important aspect in materials growth is strain. Hence, we simulate and discuss the effect of isotropic strain on the electronic and magnetic properties of cubic Fe$_2$CrTe alloy. Further, to check the effect of substrate (leading to bi-axial strain), we calculate the physical properties of the thin film (having 13 ML) of the cubic Fe$_2$CrTe alloy, supported by a suitably lattice-matched substrate MgO (probed both 5 and 7 ML). Finally, the reasonably high SP at E$_F$ (75\%) of the alloy when interfaced with MgO motivated us to perform the calculation of transmission properties of the Fe$_2$CrTe/MgO/Fe$_2$CrTe heterojunctions. We discuss and analyze the results of this heterojunction to probe its MTJ properties.

\begin{figure*}[!htbp]
\begin{center}
\includegraphics[width=1\textwidth]{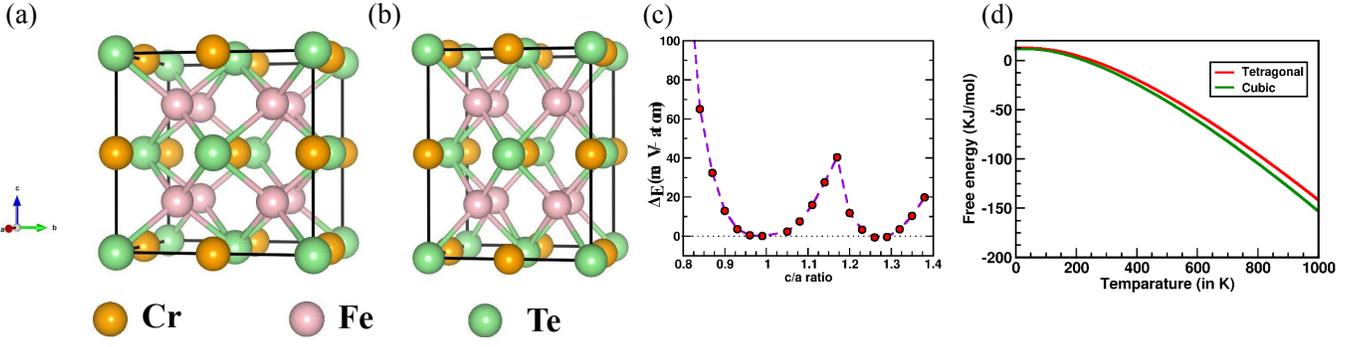}
\caption{Bulk crystal structure shown in (a) cubic and (b) tetragonal phase. (c) shows the difference in total energy (between cubic and tetragonal phases) with respect to c/a ratio for bulk Fe$_2$CrTe, where the energy for the bulk cubic phase has been taken as reference. Since we normalize the difference with respect to the energy of the cubic phase, the cubic phase (c/a = 1) corresponds to an energy value of 0 eV. (d) Shows the temperature dependent Gibbs free energy, for the cubic and tetragonal phases.}\label{Fig:1}		
\end{center}
\end{figure*}

\subsection{Bulk Physical Properties}
\subsubsection{Energetic and Dynamical Stability}
\textit{Energetics - } The binding energy(BE) of the cubic phase of Fe$_2$CrTe has been found to be -4.0252 eV per atom (for Ni$_2$MnGa the BE is -4.3963 eV per atom) and this indicates a stable alloy. As discussed above, the Fe$_2$CrTe alloy exhibits a lowest energy state in the conventional cubic Heusler alloy structure (Figure \ref{Fig:1}(a)), like the well-known Ni$_2$MnGa compound, and it has a lattice constant of 5.95 \AA. In order to assess if a stable tetragonal (martensite) phase is possible in this system or not, we vary the c/a value, by keeping the volume same as the cubic phase, since MPT is known to be a volume conserving transition. In Figure \ref{Fig:1}(b) we show a schematic figure of the tetragonal phase with the optimized c/a ratio of $\sim$1.26. In Figure \ref{Fig:1}(c), we plot the energy difference between the cubic and tetragonally distorted phase as a function of c/a. We find that the cubic (austenite) phase is very close to the energy of the tetragonal (martensite) phase (lowest energy state) showing a double-minima like plot and the energy difference is as small as $\sim$1 meV per atom. In order to cross-check this interesting observation, we carry out an all-electron calculation employing WIEN2k programme package\cite{Blaha2019WIEN2kAA} using the GGA XC term. A plot with a double-minima structure is found in this case too. We find that the energy trend is reversed but the energy difference between the two phases continues to be very small ($\sim$12 meV per atom). Similar observation of a reversal of energy ordering for very small energy difference between two phases obtained from an all-electron and a pseudopotential calculation has already been reported in the literature.\cite{Siewert_2010} Since the energy difference (between the cubic and the tetragonal phases) is very small and the difference between the phycical properties, such as total magnetic moment and density of states (DOS) are insignificant, when results from both the methods are compared, in this work, we continue to consider and present the results of physical properties, obtained from VASP.\cite{Kresse_1999,Kresse_1996}

In order to understand the energetics further and to probe the possibility of MPT, we calculate the Gibbs free energy for both the cubic and tetragonal states as a function of temperature (Figure \ref{Fig:1}(d)). Our results show that interestingly, there is no crossing of the curves, which indicates that no MPT is possible in this alloy. This would bring us to the conjecture that since the two phases are energetically rather close and no MPT seems feasible, both these phases will compete and will have equal possibility to form, depending upon the growth conditions. During growth, possibility of finding a material in two (or more) different symmetries has already been discussed by us in our group and the experimental references therein.\cite{Baral_19} However, it may be noted from Figure \ref{Fig:1}(d) that, at higher temperatures, cubic phase has a slight edge over the tetragonal phase.

\begin{figure*}[!htbp]
\begin{center}
\includegraphics[width=1.01\textwidth]{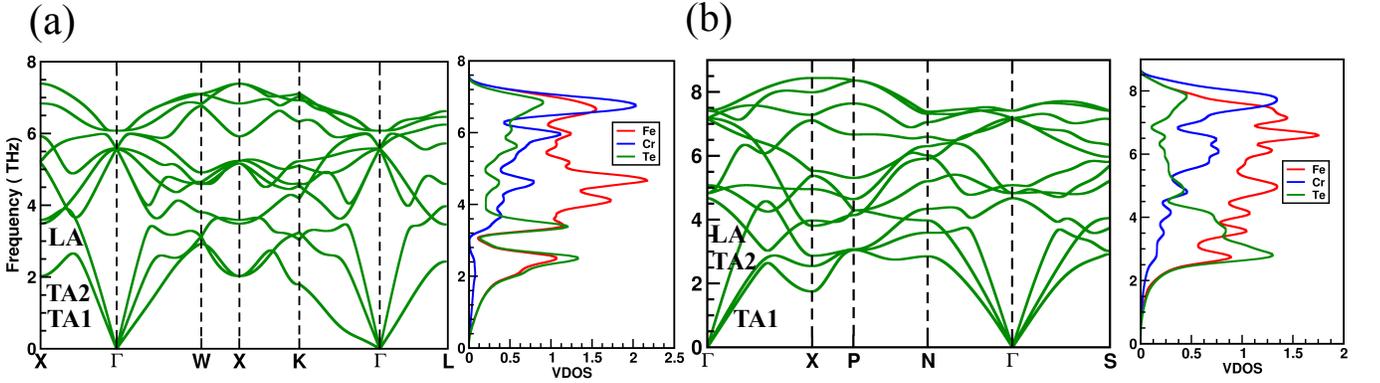}
\caption{Schematic representation of phonon dispersion relations and atom projected vibrational density of states (VDOS) of  (a) cubic and (b) tetragonal unit cell. LA, TA1 and TA2 represent the longitudinal and transverse acoustic branches, respectively.}		
\label{Fig:2}		
\end{center}
\end{figure*}

\textit{Lattice Dynamical Stability - } To probe this, we calculate the phonon dispersion curve for the cubic and tetragonal phases and the results are presented in Figure \ref{Fig:2}. For the phonon calculations, a 4$\times$4$\times$4 supercell is taken. The finite displacement method within the phonopy code\cite{Togo2015FirstPP} has been employed in order to obtain the phonon dispersion of the material. From Figure  \ref{Fig:2} it can be clearly seen that, all the frequencies are positive for both the phases, which is an important prerequisite for the lattice dynamical stability of a material. If the phonon spectra of Fe$_2$CrTe are compared with other previously studied FHAs\cite{Paul_2014,Zayak_2005}, instability in their respective cubic phases are observed. These instabilities are led by the anomalous behavior of the acoustic TA2 branch along the $\Gamma$ to X direction. However, in the present case, we do not observe clear anomalous dips in the acoustic TA2 branch in the cubic phase (Figure \ref{Fig:2}). In the cubic phase, along the $\Gamma$ to X direction, the TA1 and TA2 branches are found to be degenerate and the degeneracy is lifted in the tetragonal phase. 
The atom projected vibrational density of states (VDOS) for the cubic and tetgragonal phases (shown in Figure \ref{Fig:2} (a) and (b) ) are consistent with our discussion on the phonon spectra. As expected the Te atoms dominate the vibration at lower frequency range, whereas at the higher frequency range, Cr atoms dominate the vibration for both the phases. Hence, in Fe$_2$CrTe, in both the phases, the sequence of the optical vibration is regular, i.e. decreasing as the mass of the atom is increasing. This behavior is quite different from some other well-known FHAs, which show instability in the cubic (austenite) phase. In those systems, VDOS show unexpected anomalous behavior in the cubic phase, where optical vibration of the lighter atom is seen lying below the optical vibration of the heavier atom.\cite{Paul_2014,Zayak_2005}  These observations collectively indicate that Fe$_2$CrTe is stable in both the phases, which gets further support from the variation of the Gibbs free energy with temperature in these two phases, as discussed before.

\subsubsection{Mechanical Properties}
Having discussed the stability of the alloy, we now proceed to present and discuss the mechanical properties. For the details of the mathematical expressions for all the elastic constants and parameters, we refer to our earlier work\cite{ROY2015822} and the references therein.

\begin{table*}[ht]
\footnotesize
\begin{center}
\caption{Calculated values of elastic constants ($C_{ij}$ in GPa) for Fe$_2$CrTe in cubic and tetragonal phases. Computed values of bulk modulus ($B$ in GPa), Young's modulus ($E$ in GPa), Shear modulus ($G$ in GPa), Poisson's Ratio ($\sigma$), Cauchy pressure ($C_P$ in GPa) are also tabulated. $B$ and $G$ are calculated using the formalism given by Hill.\cite{Hill_1952}}
		\label{tab:1}
		\begin{tabular}{c c c c c c c c c c c c c c c c}
			\hline\hline
			 & Phase & $C_{11}$ & $C_{12}$ & $C_{13}$ & $C_{33}$ & $C_{44}$ & $C_{66}$ & $C^{'}$ & $B$ & $E$ & $G$ &$\sigma$ &$C_P$ &&
			 \\
			 \hline
			 & Cubic      & 208.86   & 182.96   &  -  & -        &  102.89    & -        & 13 &191.60 &130.52&47.07&0.38&80.10&& \\
			 & Tetragonal & 330.830  & 82.493   & 136.696  & 271.440  & 93.966   & 72.366   &124.17 & 182.76&228.63&88.51&0.29& -11.50&&
			 \\ \hline\hline
		\end{tabular}
	\end{center}
\end{table*}

In order to study the mechanical stability criteria, we calculate the elastic constants for Fe$_2$CrTe alloy in cubic phase. It is well-known that there are only three independent ones as $C_{11}$ = $C_{22}$ = $C_{33}$, $C_{12}$ = $C_{13}$ = $C_{23}$ and $C_{44}$ = $C_{55}$ = $C_{66}$ in the cubic phase. We list the three constants $C_{11}$, $C_{12}$ and $C_{44}$ in Table \ref{tab:1}.  It is well-known that if a cubic system fulfills the following stability criteria, then the structure is mechanically stable.\cite{Born-Huang}\\
                  $C_{11}$ - $C_{12} >$ 0 ;
                  $C_{11} + 2C_{12}>$ 0 ;
		          $C_{11}> C_{44} > $ 0 \\

In case of Fe$_2$CrTe, from the results presented in Table \ref{tab:1}, we find that all the above-mentioned criteria are satisfied, suggesting that it is a mechanically stable alloy in the cubic phase. Further, we calculate the tetragonal shear constant ($C^\prime$) and the Zener ratio (or the elastic anisotropy parameter: $A_e$) which are defined as:\\
$C^\prime$ = ($C_{11}$ - $C_{12}$)/2 \\
$A_e$ = $\frac{2\times C_{44}}{C_{11} - C_{12}}$ \\

We find that while $C^\prime$ has a value of $\sim$ 13, $A_e$ possesses a value of $\sim$8. It has been established in the literature that a negative or very small positive value of $C^\prime$ indicates an unstable cubic phase, as has been observed for Ni$_2$MnGa (a value close to 5 has been obtained both from experiments and theory)\cite{OZDEMIRKART2010177,ROY2015822} which has a non-cubic ground state.\cite{Barman_2005} The calculated value for Fe$_2$CrTe alloy can be considered to be somewhat small, as it has been seen that the typical HM systems like Co$_2$VGa\cite{Kanomata_2010}, which exhibit cubic ground state, show much higher $C^\prime$ values.\cite{Roy_2016} Further, we find that the $A_e$ value turns out to be much larger than 1. It is well-known that materials with an $A_e$ value much different from 1 often shows the tendency to deviate from the cubic symmetry and may suggest instability in the cubic phase.\cite{Luan} Presence of a minimum in energy for a tetragonal symmetry (Figure \ref{Fig:1}(c)) seems to be consistent with the low positive $C^\prime$ and large $A_e$ values corresponding to the cubic bulk phase of Fe$_2$CrTe alloy.

Now we discuss the mechanical properties, which are important and most-discussed in the literature, namely, the bulk and shear modulii, Young's modulus (E) and Poisson's ratio ($\sigma$), which are often used to describe the ductility, mallieability and overall mechanical stability of a material. Table \ref{tab:1} presents these values (all the values being rounded off up to second decimal place). For the cubic phase, we find that the values of bulk, shear and Young's modulii are somewhat higher than those of Ni$_2$MnGa.\cite{OZDEMIRKART2010177,ROY2015822} This result suggests that in case of tensile, volumetric and shear strains, the present alloy is slightly less compressible compared to Ni$_2$MnGa. In other words, it is expected to provide larger deformation resistance. However, since the $\sigma$ value is very close to that of Ni$_2$MnGa as well as many of the metals, Fe$_2$CrTe is expected to behave similar to the common metals and well-known HAs in terms of compressibility. A similar conclusion can be drawn from the calculated Pugh's ratio ($B$/$G$), which has been found to be $\sim$2.77. On an empirical level,  a material with a value higher than the critical value of 1.75 can be considered to have less inherent crystalline brittleness (ICB).\cite{S.F.Pugh} Further, according to Pettifor\cite{Pettifor}, a metal typically exhibits a high positive value of Cauchy pressure. With a value of $\sim$80, which is higher than that of Ni$_2$MnGa\cite{ROY2015822}, more metallic than directional bonding is expected in case of Fe$_2$CrTe alloy. The Kleinman parameter ($\zeta$) is a dimensionless parameter which corresponds to the relative ease of bond bending to that of bond stretching \cite{Naher2021,ma11050797}. Under given stress, bond stretching (bending) dominates if $\zeta$ is closer to 0.0 (1.0). We find a value close to 1. Hence, bond lengths are excepted to be largely unchanged if the system is distorted. 

Next we analyze the elastic stability and mechanical properties of the tetragonal phase. Table \ref{tab:1} also lists the relevant parameters for the tetragonal phase. The mechanical stability of tetragonal materials correspond to the following conditions: (1) all of $C_{11}$,  $C_{33}$, $C_{44}$, $C_{66} > $ 0; (2)($C_{11}$ - $C_{12}$) $>$ 0; (3)($C_{11}$ + $C_{33}$ - 2$C_{13}$) $>$ 0; (4) (2($C_{11}$ + $C_{12}$) + $C_{33}$ + 4$C_{13}$) $>$ 0.\cite{Born-Huang} We find all the criteria are fulfilled and hence the tetragonal phase of the material is mechanically stable. $C^\prime$ value is high and positive, indicating a stable tetragonal phase is possible from the mechanical point of view. However, larger resistance to deformation is indicated by the increased values of $B$, $G$ and $E$, compared to the cubic phase. Though the $B$/$G$ value exhibits similar value as the cubic phase, the $C_P$ value becomes negative, indicating the presence of a less metallic and more directional bonding in the tetragonal phase. 

\begin{table*}[ht]
\footnotesize
\begin{center}
	\caption{Maximum and minimum values of $E$ (in GPa), $\beta$ (in TPa$^{-1}$), $G$ (in GPa) and $\sigma$ for cubic and tetragonal phases of bulk Fe$_2$CrTe alloy.}
\label{tab:2}
\begin{tabular}{c c c c c c c c c c}
\hline
\hline
&Phase&\multicolumn{2}{c}{$E$}&\multicolumn{2}{c}{$\beta$}&\multicolumn{2}{c}{$\sigma$}&\multicolumn{2}{c}{$G$}
\\
\cline{2-10}
&& Max&Min&Max&Min&Max&Min&Max&Min
\\
\hline
& Cubic& 261.742  & 37.994 &  1.740 & 1.740 & 1.301& -0.485& 102.891& 12.950\\
& Tetragonal& 264.527 & 181.023 & 1.870 & 1.801&  0.477 & 0.052 & 124.168 & 69.220\\
\hline
\hline
\end{tabular}
\end{center}
\end{table*}

\textit{Anisotropic character of mechanical properties -} It is well-known that if the value of $A_e$ is 1, the Young's modulus turns out to be isotropic in nature. As we obtain an $A_e$ value much larger than 1, we calculate the maximum and minimum values as well as study the three-dimensional characters to probe the direction-dependent nature of various mechanical properties, including the Young's modulus. Table \ref{tab:2} gives the maximum and minimum values of Young's modulus ($E$), compressibility ($\beta$), shear modulus ($G$) and Poisson's ratio ($\sigma$). We find that except $\beta$, maximum and minimum values for rest of the parameters differ from each other. Further, we probe the directional mechanical properties in two (2D) and three (3D) dimensions. The results (presented in Figures S1 and S2\cite{Supple} for cubic and tetragonal phases, respectively) establish the anisotropic nature, more for cubic phase. 

\subsubsection{Electronic and Magnetic Properties}
\begin{figure*}[!htbp]
	\begin{center}
		\includegraphics[width=1\textwidth]{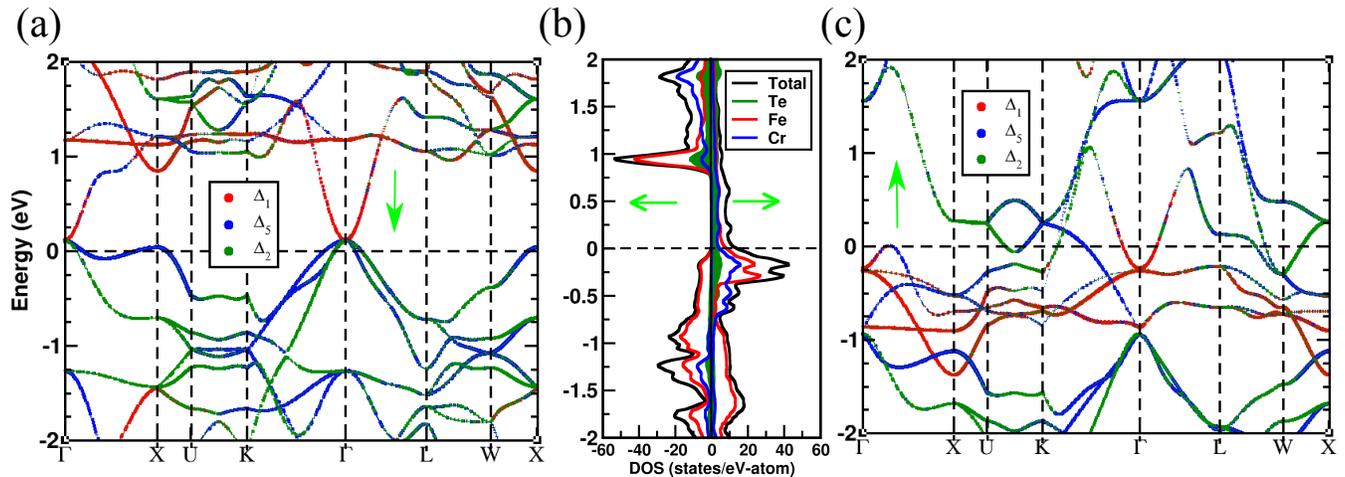}
		\caption{Bulk electronic structure of cubic Fe$_2$CrTe: (a) and (c) depict orbital projected band structure of minority and majority electrons and (b) gives atom projected density of states. Here $\Delta_1(s,p_z,  d_{z^2})$, $\Delta_2(d_{xy},d_{x^2 -y^2})$ and $\Delta_5(p_x, p_y, d_{xz}, d_{yz})$ represent the orbital symmetries of the bands.} 		
		\label{Fig:3}		
	\end{center}
\end{figure*}

\begin{figure*}[!htbp]
	\begin{center}
		\includegraphics[width=1\textwidth]{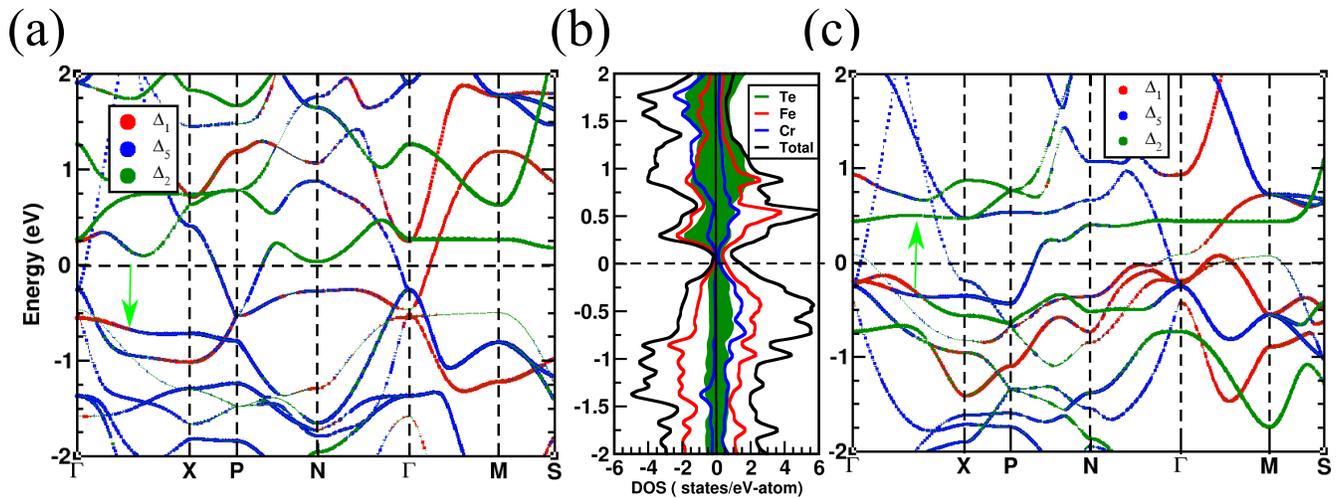}
		\caption{Bulk electronic structure of tetragonal Fe$_2$CrTe: (a) and (c) show orbital projected band structure of minority and majority electrons and (b) presents atom projected density of states. Here $\Delta_1(s,p_z, d_{z^2})$ , $\Delta_2(d_{xy},d_{x^2 -y^2})$ and $\Delta_5(p_x, p_y, d_{xz}, d_{yz})$ represent the orbital symmetries of the bands.}		
		\label{Fig:4}		
	\end{center}
\end{figure*}

\textit{DOS and Band Structure - }In Figure \ref{Fig:3}, we plot the density of states (DOS) and the band structures of the cubic phase. From the total density of states, we find that there is a high SP at E$_F$. It is found to be about 95\%. The E$_F$ for the equilibrium lattice constant is found to be located on the valence band edge. The majority spin DOS in the occupied region near E$_F$ has dominant contribution from the Cr and Fe d states, with a small contribution from the Te atoms (Figure \ref{Fig:3}(b)). On the contrary, in case of the minority spin, Fe DOS contributes much more near and at the E$_F$ than the other atoms. These Fe d states in the minority spin channel leads to the reduction of the SP at E$_F$ from 100 to 95\%. Further, the unoccupied states in the conduction bands near E$_F$ also has dominant Fe-d character. The orbital projected band structure of Fe$_2$CrTe in the cubic phase is shown in Figure \ref{Fig:3}(a), (c). The $\Delta_1$ symmetry is associated with the $s, p_z, d_{z^2}$ orbital character. In contrast, $p_x, p_y, d_{xz}, d_ {yz}$ orbitals specify $\Delta_5$ symmetry and $d_{xy}, d _{x^2 −y^2}$ orbitals are assigned to the $\Delta_2$ symmetry.  We can see the presence of highly disperssive parabolic electron like conduction band in the both majority and minority spin channels around the $\Gamma$ point and these bands have dominant contribution from $\Delta_1$ symmetric bands, originating from Te 5s states. However, we do not observe any bands crossing the E$_F$ along the $\Gamma$ to X ($i.e.$ along $<001>$) direction in the majority spin channel (Figure \ref{Fig:3}(c)), which has a great importance in the spin dependent transport properties and we discuss it in the later part of our paper. However, in the minority spin channel, we find two bands cross the E$_F$, which have dominant $\Delta_5$ and $\Delta_2$ orbital symmetries. Further, in the minority spin case the top of the valence band just touches the E$_F$, leading to negligible DOS at E$_F$.

In case of the tetragonal phase (Figure \ref{Fig:4}), while the majority DOS contributions near the F$_F$ remain similar (both from Fe and Cr d states), the minority DOS at (and also above) E$_F$ gets slightly populated by both Fe and Cr d states, Fe states having more contribution. Due to the increased DOS at E$_F$ for the minority channel, the SP at E$_F$ gets reduced significantly (67\%). Other than at the E$_F$, overall DOS of the tetragonal phase for the majority and minority spin channels also show that the peak positions are rather close to each other unlike the cubic case. This leads to a much lower magnetic moment for the former phase. Further, the orbital projected band structure in Figure \ref{Fig:4}(a), (c) suggests that parabolic conduction and valence bands around the $\Gamma$ point in the minority spin channel is pushed away from the E$_F$ and move towards the higher (lower) energy side for the conduction (valence) bands when compared to the cubic case. From the symmetry analyses, we confirm the presence of $\Delta_1$ symmetric band along the $\Gamma$ - M direction ($i.e$ along the $<001>$ direction), in both the spin channels. We have further found that there are spaghetti of bands around the high symmetry point $\Gamma$ in both the spin channels, increasing the valley degeneracy at $\Gamma$. 

\begin{figure*}[!htbp]
	\begin{center}
		\includegraphics[width=.6\textwidth]{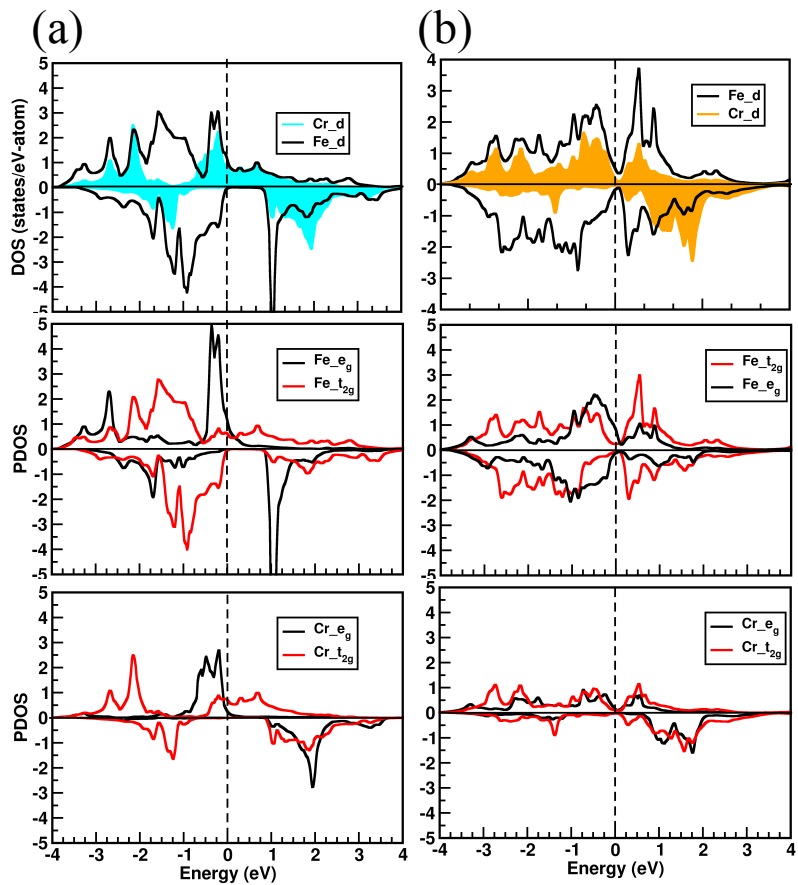}
		\caption{Atom and orbital projected DOS (PDOS) of Fe$_2$CrTe: (a) cubic, (b) tetragonal Phase, respectively. The e$_g$ states have d$_{x^2 -y^2}$ and d$_{z^2}$ orbital contributions and the t$_{2g}$ states have d$_{xy}$, d$_{xz}$ and d$_{yz}$ orbital contributions. }		
		\label{Fig:5}		
	\end{center}
\end{figure*}

Figure \ref{Fig:5} shows the DOS contribution of the various d states of the transition metal elements (Fe and Cr) as these populate the DOS near and at E$_F$. The splitting of the e$_g$ and t$_{2g}$ like states in the cubic case, specifically in case of Cr, which is in an ocatehedral symmetry, being surrounded by Fe atoms, is clearly visible from Figure \ref{Fig:5} (bottom panel). Although the Fe atoms are the 2nd nearest neighbor of other Fe atoms, the hybridization between them is qualitatively more important. The t$_{2g}$ and e$_g$ like states for Fe atoms, which is in a tetrahedral symmetry, due to the four transition metal atoms Cr as neighbor can be seen from Figure \ref{Fig:5} (middle panel). This aspect of crystal-field splitting of the t$_{2g}$ and e$_g$ like states has been shown to play an important role in yielding a HM like behavior in cubic half and full Heusler alloys.\cite{Baral_19, Galanakis} The DOS of Cr atom shows much larger energy gap  ($i.e.$ larger t$_{2g}$ and e$_g$ splitting) around the E$_F$, larger than what has been observed for Fe$_2$CrTe bulk (Figure \ref{Fig:5} (top panel)). However, the real gap is determined by the Fe-Fe interaction and the t$_{2g}$, e$_{g}$ splitting of the Fe atoms. On the contrary, for the case of tetragonal symmetry, due to the absence of clear octahedral/tetrahedral symmetric geometrical environment for the magnetic elements, and consequent absence of crystal-field effect, the splitting between t$_{2g}$ and e$_g$ like states is not clear (Figure \ref{Fig:5} (b)). This might have led to the smaller SP value at E$_F$ as compared to the cubic case. 

\begin{figure*}[!htbp]
	\begin{center}
		\includegraphics[width=.6\textwidth]{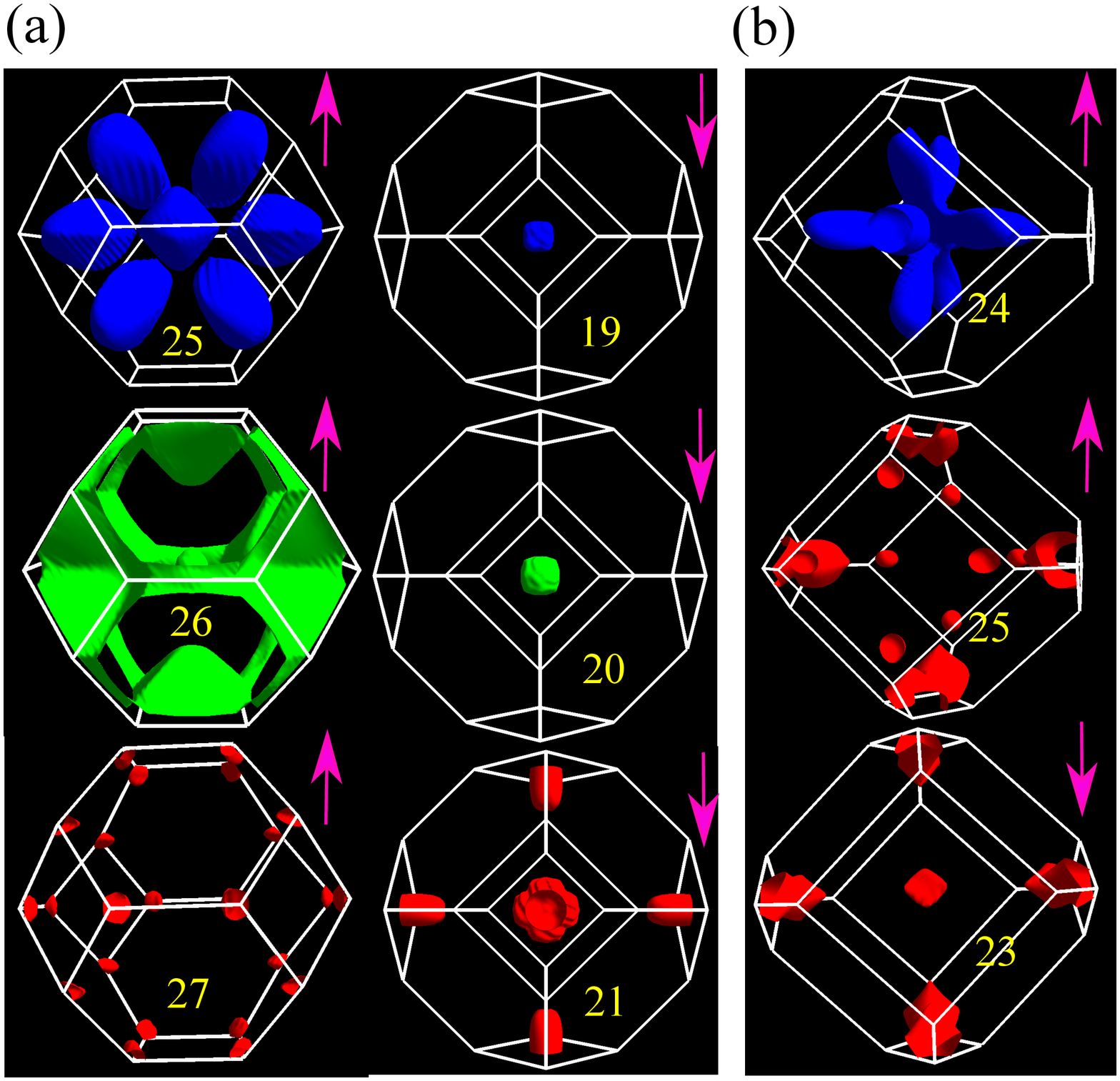}
		\caption{Spin-polarized Fermi surface of Fe$_2$CrTe: (a) and (b) panels correspond to the cubic and tetragonal phases, respectively. The $\uparrow$ and $\downarrow$ represent Fermi surface due to majority and minority electrons, respectively. We have also mentioned the band index of the bands crossing the E$_F$, which constitute the Fermi surfaces and those particular bands are shown in Figures S6 and S7\cite{Supple} for the cubic and tetragonal phase, respectively. }		
		\label{Fig:6}		
	\end{center}
\end{figure*}

	\textit{Effect of Hubbard U term -} We have also addressed the role of onsite Coulomb interaction (Hubbard U) on the electronic and magnetic properties of Fe$_2$CrTe in both the phases. The electron - electron Coulomb interaction and the self-interaction correction are considered in the rotationally invariant way (GGA+U) according to the Dudarev’s method \cite{Dudarev}. We have considered the U value of Fe and Cr to be 3 eV and 2 eV, respectively. This is in accordance with previous studies reported in the literature.\cite{BNCox_1974,Paudel,CHAKRABARTI20123547}. The results are found to be significantly different for both the phases. In the cubic phase the SP at E$_F$ is found to be changed drastically with the Hubbard U parameter (Table S1).\cite{Supple} Further analyses on atom-projected DOS indicate that electronic and magnetic properties show significant dependence on U$_{Fe}$ and minimal dependence on U$_{Cr}$(Figure S3).\cite{Supple} However, the SP value in the tetragonal phase shows moderate change over the range of U$_{Fe}$ and U$_{Cr}$ values, considered in our calculation (Table S1).\cite{Supple} and the electronic structure is found to be less affected as compared to the cubic phase (Figure S3).\cite{Supple} Usually the strength of Hubbard U for each atom in different local environments can be easily estimated by seeking a good agreement between the calculated and the experimental results. However, due to the predictive nature of the present work, the present results await experimental validation.

\textit{Fermi Surface - } We present the results of calculated Fermi surfaces(FS) for the cubic and tetragonal phases of Fe$_2$CrTe alloy, for both majority and minority electrons in Figure \ref{Fig:6}. Further, bands with the respective band indices are shown in Figures S6 and S7\cite{Supple}, whereas Figure.S5 shows the positions of the high symmetry k-points in the irreducible Brillouin zone.\cite{Supple} By analyzing these figures, we observe the following for the cubic phase. The minority spin channel has three bands, 19, 20 and 21, which are mostly Fe derived (small contributions from the Te atoms) and cross E$_F$. While the former two bands give rise to only a carrier pocket at the $\Gamma$ point, band number 21 additionally shows a pocket at the X point. On the other hand, in the majority spin channel the bands, 25, 26 and 27 are mostly Cr derived and have small contributions from Fe and Te atoms, specially for band 26. The FS due to band 26 forms a open spherical cone, indicating an electron-like behavior, whereas the FS is hole-like for band 25. Apart from that, small electron-like pockets are also observed at the W point, due to band 27. 

In the tetragonal phase, the character of the minority FS is shared by both Fe d and Cr d electrons (see Figure \ref{Fig:4}). The minority spins generate very small electron-like pockets at the X point and hole-like pockets at the $\Gamma$ point. But the majority spin FS undergoes significant changes as we go from cubic to tetragonal phase. The majority FS due to band 25, produces small electron-like pockets at the high symmetry points X and N. Large hole-like pockets, centered around the $\Gamma$ point, can be seen due to band 24. This drastic change in the majority spin FS (as we go from cubic to tetragonal phase) is also accompanied by a change in the spin magnetic moment, as seen for the Fe atoms, which we will discuss in the next section.

\begin{table*}[ht]
	\footnotesize
	\begin{center}	
\caption{Calculated values of total and atom-projected moments (in $\mu_B$) for both the cubic and tetragonal phases.}
\label{tab:3}
\begin{tabular}{c c c c c c c}
\hline
\hline
Phase & &$\mu_T$/f.u. & $\mu_{Fe}$ & $\mu_{Cr}$ & $\mu_{Te}$&
\\
\hline
\textbf{Cubic}	&GGA &3.99  &0.89 & 2.22& -0.003&
\\
\textbf{Tetragonal} & GGA&1.99 &-0.23 & 2.31&  0.004&
\\
\hline
\hline
\end{tabular}
\end{center}
\end{table*}

\begin{figure*}[!htbp]
	\begin{center}
		\includegraphics[width=.8\textwidth]{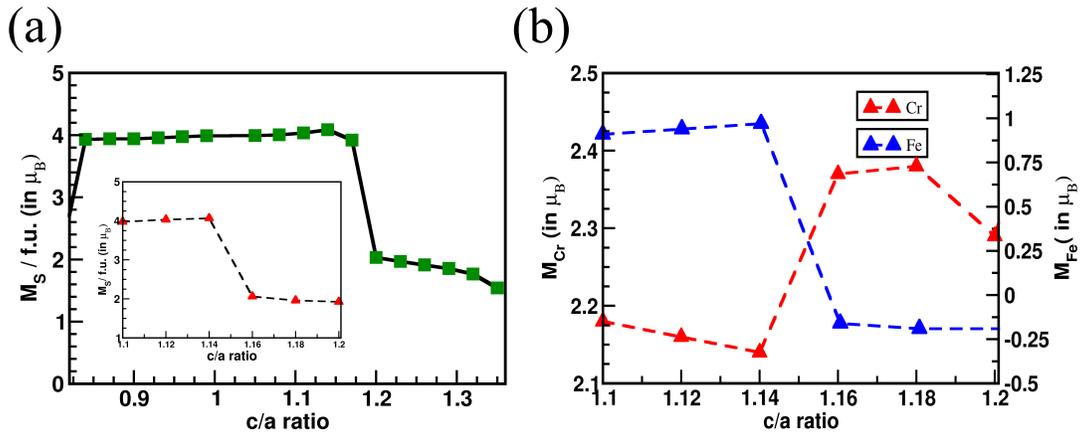}
		\caption{(a) and (b) exhibit variation of total and partial magnetic moments, respectively, plotted as a function of c/a.}		
		\label{Fig:7}		
	\end{center}
\end{figure*}

\textit{Magnetic Properties - } As it is clear from Table \ref{tab:3}, the total magnetic moment of the system significantly reduces in the tetragonal phase. This happens due to the change in the partial magnetic moment of the Fe atom. From an overall ferromagnetic coupling observed in the cubic phase, due to the negative moment of Fe, the system assumes an overall ferrimagnetic configuration in the tetragonal phase.  The partial moment of the Cr atom remains quite similar in both the cubic and non-cubic cases. The spin-polarized DOS of the d states of the Fe and Cr atom (Figure \ref{Fig:5}) give clear indication of these. To explore this in more detail, in Figure \ref{Fig:7}, we plot the total and partial magnetic moments of both the atoms, as a function of c/a values. It is observed that at about  c/a of $\sim$1.15, a transition is observed for both the total and partial atomic moments. While the  partial moment value of Fe goes from positive ($\sim$1 $\mu_B$) to a negative value (-0.23 $\mu_B$), partial moment of Cr atom make a transition of value of about 2.1 to close to 2.4 $\mu_B$. Hence, it is clear that though Cr is the main moment-carrying atom in both the phases, it is the partial moment of Fe atom which leads to a change in magnetic configuration of the system. Additionally, it appears that such a transition of magnetic configuration from ferromagnetic in the cubic phase to ferrimagnetic in the tetragonal phase is independent of the Hubbard U parameter used in this study (Figure S4).\cite{Supple} 

\begin{figure*}[!htbp]
	\begin{center}
		\includegraphics[width=1\textwidth]{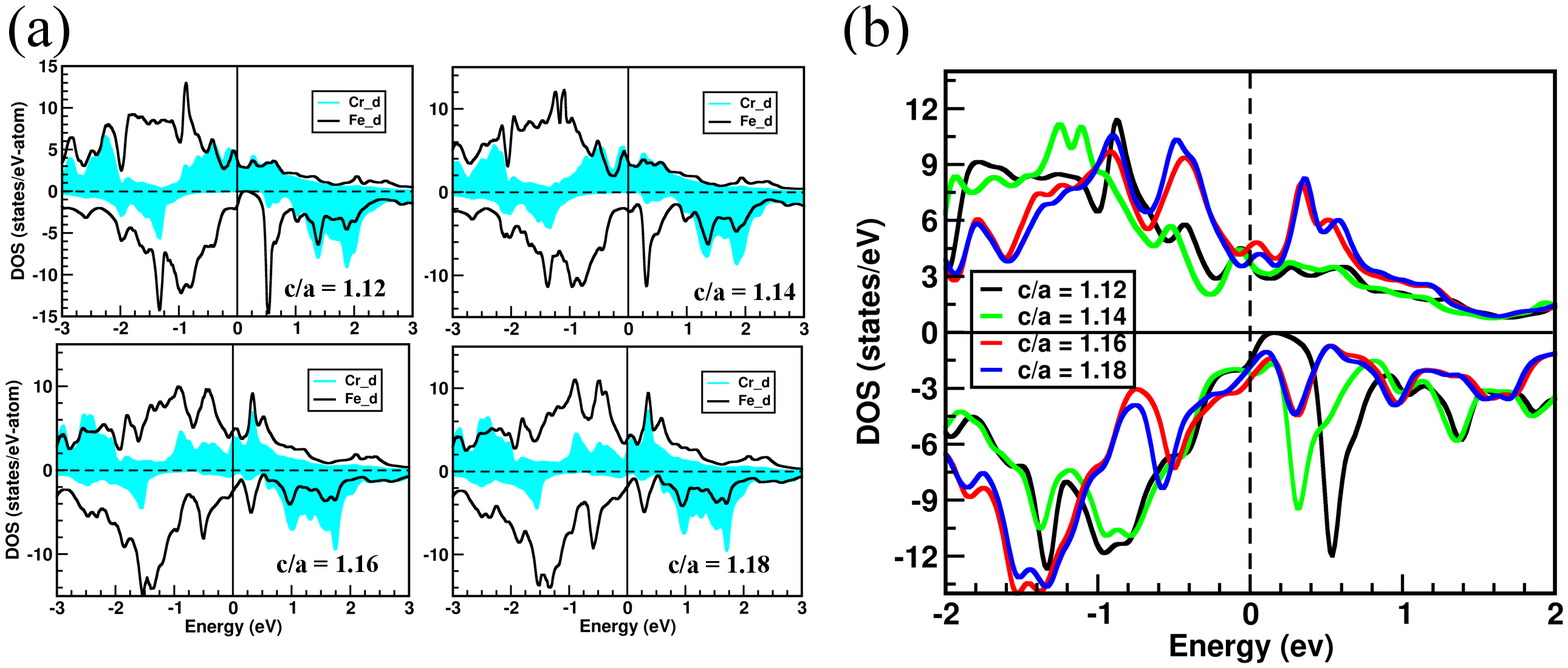}
		\caption{Spin-polarized density of states for tetragonal distortion in Fe$_2$CrTe for different c/a values. (a) presents orbital projected density of states for both the Cr and Fe atoms and (b) shows the DOS of d orbitals of the Fe atoms corresponding to minority and majority electrons, respectively.}		
	\label{Fig:8}		
	\end{center}
\end{figure*}

In order to understand the drastic change in the magnetic properties at a particular value of c/a, in Figure \ref{Fig:8}, we plot the spin-polarized DOS for different c/a values around the value of c/a = 1.15. We find that the Cr atom projected DOS (PDOS) is less affected than the Fe PDOS (Figure \ref{Fig:8} (a)). But the overall intensity of the minority spin DOS for Fe seems to have increased. Further, we have observed strong hybridization between the majority spin states of Fe and Cr d electrons near E$_F$ (-0.50 to 0 eV) for the c/a values 1.12 and 1.14, which is absent for the c/a values beyond $\sim$1.15 (Figure \ref{Fig:8}(a)). This might give rise to to the difference in Fe and Cr spin moments  below and above c/a = 1.15. In Figure \ref{Fig:8}(b), we have shown the DOS of the Fe-d states for different c/a values near the E$_F$. The peak around $\sim$-0.10 eV in the majority DOS is shifted to $\sim$0.10 eV as we increase the c/a ratio. Further the peak around $\sim$ -1 eV in the majority DOS is also pushed to the higher energy ($i.e.$ towards E$_F$) as c/a ratio changes from 1.12, 1.14 to 1.16 and 1.18. On the contrary in the minority spin states, we see the trend is quite opposite. The minority DOS (around $\sim$ 0.5 eV and -1 eV) are pushed from the lower binding energy to the occupied side with increasing c/a ratio. This is also corroborates well with decrease in exchange-splitting energy of the Fe atoms from 0.91 eV to 0.07 eV as c/a ratio changes from 1.14 to 1.16. On the other hand for the Cr atom the change in the exchange-splitting energy is rather small ( 1.88 eV and 1.82 eV for c/a ratio 1.14 and 1.16 respectively). The exchange-splitting energy is obtained by calculating the  d-band centers of the atoms using VASPKIT programme \cite{WANG2021108033}.

\subsubsection{Effect of Isotropic Strain on the Electronic, Magnetic and Elastic Properties of Cubic Fe$_2$CrTe}
\begin{figure*}[!htbp]
	\begin{center}
		\includegraphics[width=1.\textwidth]{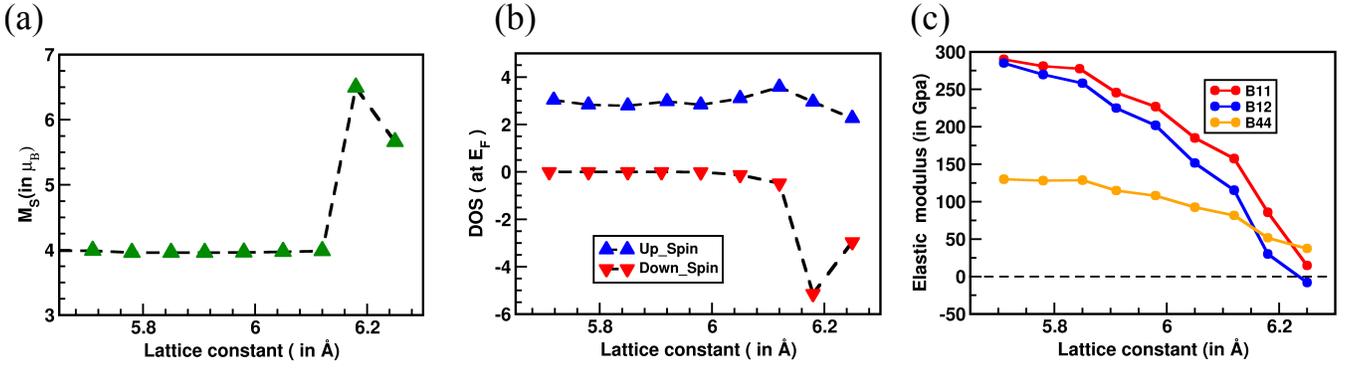}
		\caption{Effect of strain (isotropic pressure) on the (a) magnetic properties; (b) electronic properties; (c) elastic properties of Fe$_2$CrTe in the cubic phase.}	
		\label{Fig:9}		
	\end{center}
\end{figure*}

It is well-known that, any change in the geometrical properties of Heusler alloy system may affect its electronic, magnetic and mechanical properties\cite {Zhang,SINGH2013201,RAY2021158906}. Hence, it is important to probe the effect of lattice constant variation on the electronic structure and magnetic properties of the bulk material. One of the simplest ways is to apply an external pressure or equivalently apply a uniform strain in the system. In our case, we have investigated the electronic, magnetic properties and mechanical stability of the system by applying a uniform strain of $\pm 5 \%$ (changing pressure from -20 GPa to 20 GPa). In Figure \ref{Fig:9}, we have shown these effects. The total magnetic moment of the system remains nearly integer ($\approx 3.99 ~\mu_B$), almost over the entire range of latice constants (5.71 - 6.10 \AA) and after that the total magnetic moment suddenly increases at 6.22 \AA and thereafter decreases (Figure \ref{Fig:9}(a)). Further, we do not observe any magnetic phase transition ( ferromagnetic to anti-ferromagnetic) over the studied range of lattice constant. Since the cubic phase of Fe$_2$CrTe shows high spin polarization, we have also verified its HM property under uniform strain (Figure \ref{Fig:9}(b)) and found that it shows close to 100 $\%$ SP at E$_F$ on applying a negative uniform strain in the system. We further study the mechanical stability of the system, when the lattice constant changes. Under uniform pressure (P), the elastic constants are modified according to the following equation\cite{Zhang}:\\

\begin{center}
	$B_{11} = C_{11} - P ~;~ B_{12} = C_{11} +P~ ;~ B_{44} = C_{44} -P~$ 
	\end{center}
	
	\begin{center}
and the stability criteria changes to:
	$B_{11} - B_{12} > 0 ~; B_{11} + 2B_{12} >0~; B_{44}>0$ 
\end{center}

In Figure \ref{Fig:9}(c), we have shown the change of elastic constants with the lattice constant under uniform strain.  The compound is found to be mechanically stable over almost the whole range of lattice constants studied, as it satisfies the stability criteria as discussed above. However, when the lattice constant is too large ($\sim$6.25 \AA), the compound is no longer mechanically stable.This is due to the fact that with increasing lattice constant, the interaction between the atoms weakens and the stability is thus destroyed.

\subsection{Fe$_2$CrTe on MgO Surface}
From our previous discussion, we have established that the cubic phase of Fe$_2$CrTe behaves like a nearly HM system. But in reality, HM properties can be highly affected by any kind of crystal disorders; such as defects, surfaces and interfaces. The study of a HM system on a substrate and corresponding surface and interface-related effects on its electronic and magnetic properties can lead to interesting results. In order to simulate an interface, we have constructed the Fe$_2$CrTe/MgO(001) system (Figure S8)\cite{Supple}, by placing the O atoms (a) on top of the Cr and Te atoms $i.e$ on a Cr-Te terminated interface and (b) on top of the Fe atoms of a Fe-Fe terminated interface. The Cr-Te terminated interface is found to be energetically more stable. This corroborates well with previous studies on Heusler alloy and MgO based heterojunctions\cite{Miura_08,Peter_2009,Saito_2010}, where it has already been reported that YZ interface of X$_2$YZ FHA, where Y and Z atoms are situated on the top of O atoms, is the most stable one. Hence we consider this interface for further studies. Here we have considered 13 mono-layers (ML) of Fe$_2$CrTe and 7 ML of MgO with $sim$15 \AA~ of vacuum to prevent interaction between the adjacent surfaces in a periodic arrangement. Further, the in-plane lattice constant of the Fe$_2$CrTe/MgO surface, was fixed at 4.21 \AA ~($\frac{a}{\sqrt{2}}$, a being the lattice constant of bulk Fe$_2$CrTe) in the cubic phase, which has an excellent lattice matching with bulk MgO (4.21 \AA).

\begin{figure*}[!htbp]
	\begin{center}
		\includegraphics[width=.7\textwidth]{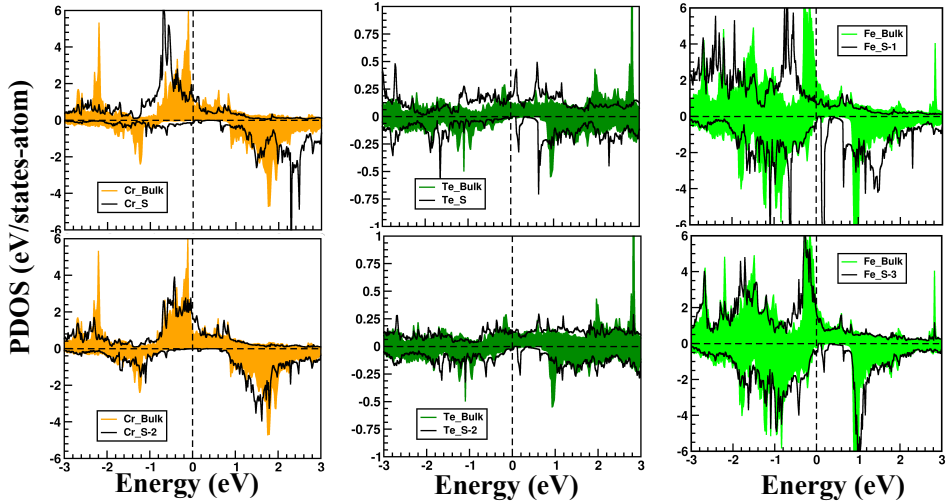}
		\caption{The spin-polarized, atom projected density of states are shown for atoms at different layers for the composite system with Cr-Te/MgO interface. Here S, S-1, S-2, and S-3 represent surface and subsurface layers as we go away from the interface, respectively.}		
		\label{Fig:10}		
	\end{center}
\end{figure*}

First we discuss the stability of this interface in the $ab-initio$ DFT framework. We have calculated the binding energy and also the surface free energy ($\gamma$) \cite{Zarei_2008}, which is defined as,\\
\begin{center}
 $\gamma =\frac{G(T,P_i)-\sum_{i}{n_i\mu_i}}{2A} $
\end{center} 

where, G is the Gibbs free energy of the surface, $n_i$ and $\mu_i$ are the number and chemical potential of the i$^{th}$ element and A is the surface area of the supercell. The binding energy and surface free energy ($\gamma$) are $\sim$ -4.3767 eV/atom and -1.9781 ev/\AA$^2$,~ respectively and these indicate a stable composite system. From Table \ref{tab:4}, we see that there is some buckling  in the interface and subsurface Cr-Te layers, where the Te atoms move towards the substrate (MgO) side due to higher electronegativity of Te as compared to Cr. The increase of magnetic moments of the interface atoms (Table \ref{tab:4}) is a well-known phenomenon\cite{Joydipto_21,Habibi_2013}, when the lower hybridization at the surface leads to the enhancement in exchange-splitting of the interface atoms. This can also be confirmed from the atom projected spin-polarized DOS (Figure \ref{Fig:10}), where we see the majority (minority) spin states of the Cr atoms shift towards higher (lower) binding energy with respect to E$_F$, as compared to bulk Cr. It is also evident from Figure \ref{Fig:10}, that the interface states of the minority spin channel are mostly localized at the Fe atoms of the sub-surface layer. As a result, the SP of the surface is significantly reduced as compared to bulk. However, the surface effect is able to penetrate a few ML and rest of the layers show bulk-like properties (as is evident from Table \ref{tab:4} and Figure \ref{Fig:10}). 
 
\begin{table*}[ht]
	\footnotesize
	\begin{center}
		\caption{Calculated electronic, geometric and magnetic properties of Fe$_2$CrTe/MgO interface. Surface buckling ($\delta$l in \AA), atomic magnetic moments (in $\mu_B$) and the SP (in $\%$) at E$_F$  are shown in the table. Bulk values are also given for comparison.}
		\label{tab:4}
		\begin{tabular}{c c c c c c c c  }
			\hline\hline
			&  Atomic Layer     & $\delta$l & SP &  \multicolumn{2}{c}{Atomic Magnetic Moments} &  &\\
			\hline\hline
			&     &    &    &  Cr & Fe &Te  &      \\
			\hline
			& Interface (S) & 0.21  &   71 & 3.12  &  -- & -0.05  &       \\
				& S-1 & 0.00 &   11 & --  & 1.82  &  --&        \\
					& S-2 & 0.20 &  94 & 2.05  &   &-0.07&          \\
					& S-3 & 0.00 &  41 &--   &  0.98 &     --&     \\	
				& S-4 & 0.00 &  94 &  2.10 & -- &-0.05&
				\\
				& Bulk & -- & 95& 2.22 &  0.89  &	-0.04&
				\\
			 \hline\hline
		\end{tabular}
	\end{center}
\end{table*}

\subsection{Spin-Transport Properties of Fe$_2$CrTe/MgO/Fe$_2$CrTe Heterojunction}

\begin{figure*}[!htbp]
	\begin{center}
		\includegraphics[width=.7\textwidth]{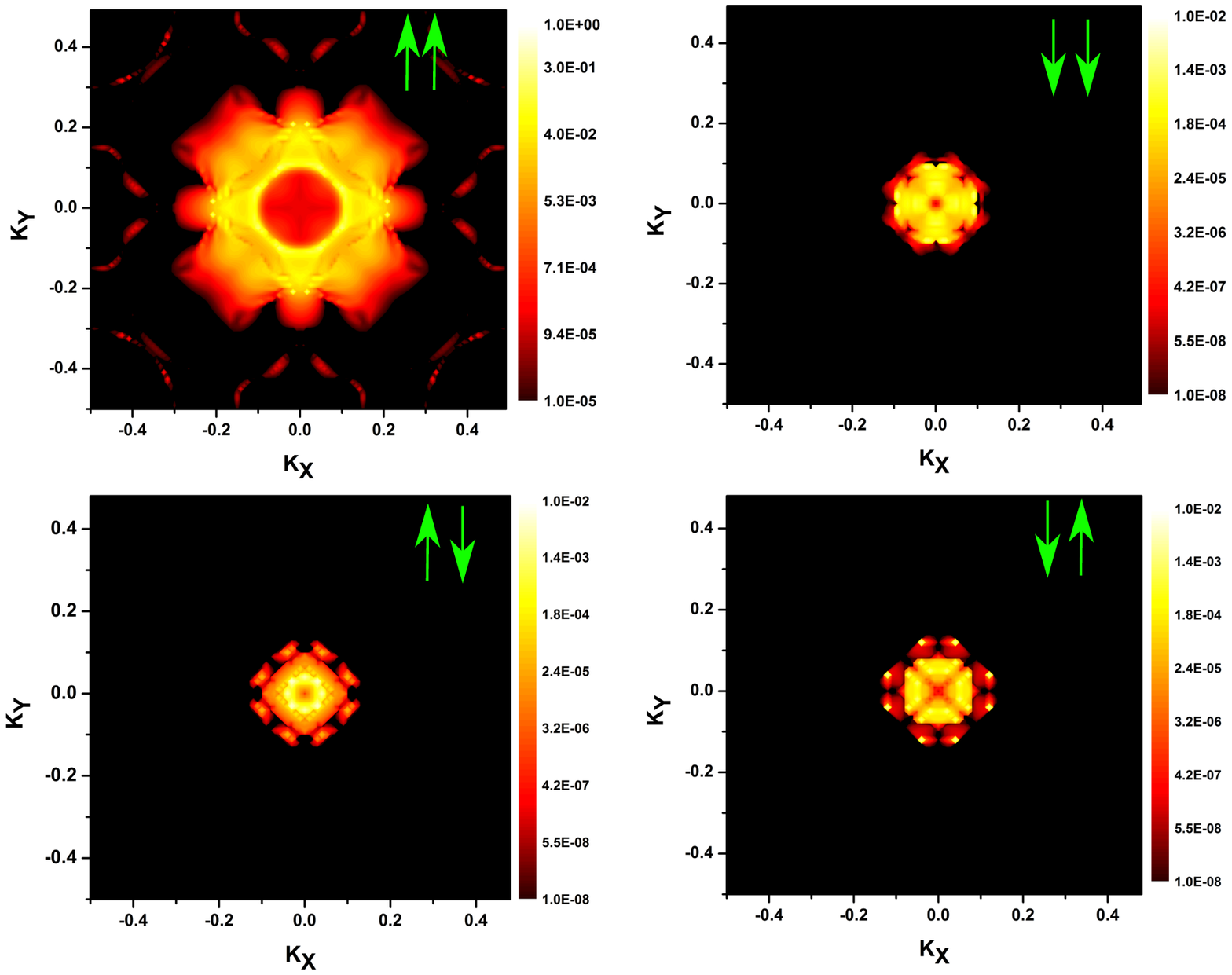}
		\caption{K-resolved Transmittance in parallel magnetization case for Fe$_2$CrTe/MgO/Fe$_2$CrTe heterojunction with Cr-Te interface for 5 ML of  with cubic phase of Fe$_2$CrTe: Top and bottom panels are for parallel (PC) and anti-parallel spin configurations (APC), respectively.}		
		\label{Fig:11}		
	\end{center}
\end{figure*}

\begin{table*}[ht]
	\footnotesize
	\begin{center}
		\caption{Calculated transport properties of Fe$_2$CrTe/MgO/Fe$_2$CrTe junctions  for the cubic and tetragonal phases.  G$^{\uparrow\uparrow}_{PC}$ (G$^{\downarrow\downarrow}_{PC}$) are spin-up (spin-down) conductance at the Fermi energy for parallel (P) spin configuration. G$^{\uparrow\downarrow}_{PC}$ (G$^{\downarrow\uparrow}_{PC}$) are the spin up-to-spin down and spin down-to-spin up conductance in the anti-parallel configuration state (APC), respectively. The tunnel magnetoresistance ratio is defined as $\frac{G_{PC} -G_{APC}}{G_{APC}}$, where G$_{}P = G^{\uparrow\uparrow}_{PC} +  G^{\downarrow\downarrow}_{PC}$ and G$_{APC} = G^{\uparrow\downarrow}_{APC} +  G^{\uparrow\downarrow}_{APC}$.} 
		\label{tab:5}
		\begin{tabular}{c c c c c c c c c}
			\hline\hline
			&Phase& Layer thickness & G$^{\uparrow\uparrow}_{PC}$& G$^{\downarrow\downarrow}_{PC}$ &  G$^{\uparrow\downarrow}_{APC} $& G$^{\downarrow\uparrow}_{APC}$&TMR ($\%$) &\\
			\hline\hline
		&Cubic	&   5  & 3.567$\times10^{-3}$ &  5.342$\times10^{-6}$  &  6.678$\times10^{-6}$&   
			  5.372$\times10^{-6}$  & 300&   \\ 
		
			&Tetragonal & 5&  7.050$\times10^{-3}$  &  4.461$\times10^{-5}$&   
			  3.048$\times10^{-7}$&1.716$\times10^{-6}$  & 3522&   \\
		\end{tabular}
	\end{center}
\end{table*}

\begin{figure*}[!htbp]
	\begin{center}
		\includegraphics[width=.7\textwidth]{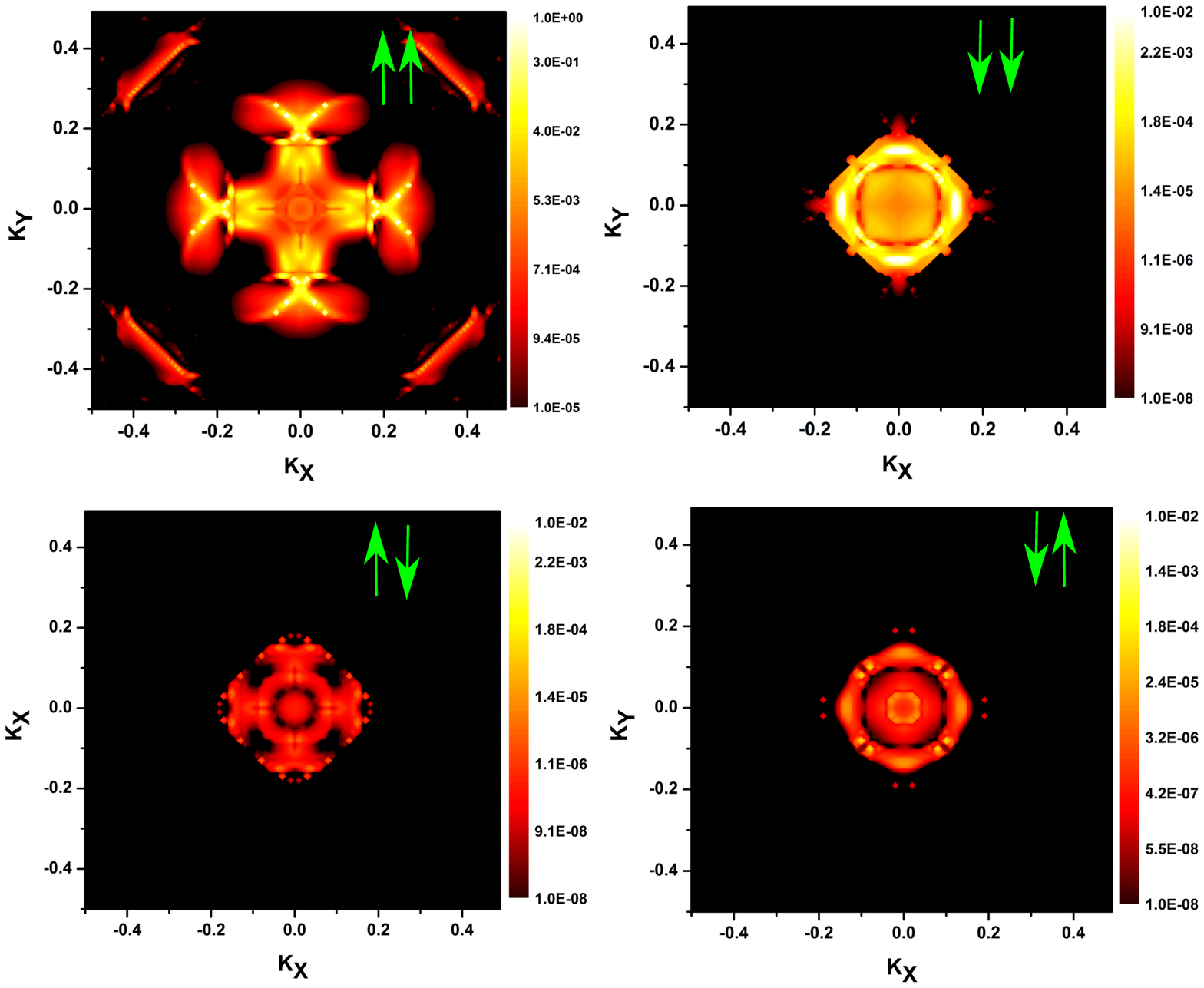}
		\caption{K-resolved Transmittance in parallel magnetization case for Fe$_2$CrTe/MgO/Fe$_2$CrTe heterojunction with Cr-Te interface for 5 ML of MgO and with tetragonal phase of Fe$_2$CrTe: Top and bottom panels are for parallel (PC) and anti-parallel spin configurations (APC), respectively.}
		\label{Fig:12}		
	\end{center}
\end{figure*}

From our previous discussion, we predict that the Fe$_2$CrTe cubic alloy may be grown on MgO(001) substrate as it forms an energetically stable surface. In bulk Fe$_2$CrTe the half-metallic behavior is affected due the highly disperssive bands in the minority spin channel due to the dominant contribution from the Fe atoms (Figure \ref{Fig:3}), which is also evident from the non-integer spin magnetic moment ($\sim 4$ $\mu_B$). Thus transport in both the spin channels is expected in case of cubic phase of Fe$_2$CrTe. Further, the SP at E$_F$ for the tetragonal phase has been found to be much less than that of the cubic phase. However, we find a significantly strong SP of transport even at high temperature, when we calculate the temperature dependence of the spin-polarized conductivity (Figure S9)\cite{Supple}, specially in the cubic phase. This indicates even though Fe$_2$CrTe may not completely be half-metallic, it may be a useful for a spin-injector material even at higher temperature. 

 MTJ materials with an electrode with a reasonably high SP at E$_F$ have recently gained lots of attention, due to their high TMR ratio, where the resulting current in the junction strongly depends on the relative magnetization of the electrode.\cite{Peter_2009,Miura_2008} Here we investigate the Fe$_2$CrTe/MgO/Fe$_2$CrTe heterojunction to explore its  MTJ properties, in both the phases.  It may be noted that the tetragonal phase of Fe$_2$CrTe exhibits a lattice mismatch of 7$\%$ with MgO (in-plane lattice constant = a/$\sqrt{2}$).  Here we have constructed our heterojunction with 13 ML of Fe$_2$CrTe and have taken 5 ML of MgO. In Table \ref{tab:5}, we have shown the transmittance value and TMR ratio of the heterojunction for cubic and tetragonal phases of Fe$_2$CrTe, where the TMR value is defined as  $\frac{G_{PC} -G_{APC}}{G_{APC}}$, where G$_{PC}$ and G$_{APC}$ indicate the total conductance for the MTJs in the parallel (PC) and anti-parallel (APC) state. Despite the fact that the cubic phase has a higher SP value, the TMR ratio indicates a 10 fold rise in the TMR ratio value for the tetragonal phase compared to the cubic phase (Table \ref{tab:5}). To understand this, we examine the majority spin band structure of both the phases along the $<001>$ direction (Figure  \ref{Fig:3} and Figure  \ref{Fig:4}).  It has already been established that the $\Delta_1$ band in the majority spin state is essential to obtain a large TMR ratio for MgO-based MTJs.\cite{Miura_2008,Butler_2008} We see that the bands with $\Delta_1$ symmetry along the $<001>$ direction are only present for the tetragonal phase. Here we must keep in mind the possibility that these theoretical values can be constrained by the presence of a variety of disorders and defects at the interface that may develop during the sample growth. Therefore, our predicted values of the TMR ratio define the upper limit of the same.

For a comprehensive understanding of the above spin-dependent conductance and TMR effect, in Figure \ref{Fig:11} we show the k-resolved  transport properties at the E$_F$ of the heterojunction with MgO (with layer thickness of 5 ML) in PC and APC state for the cubic phase of Fe$_2$CrTe. The majority spin transmission in PC state is found to be negligible around the center of the BZ, though MgO has a small decay constant ($\kappa$) around $\Gamma$ point, as can been seen from the complex band structure of MgO (Figure S10).\cite{Supple} It is evident from the orbital projected majority spin band structure of bulk Fe$_2$CrTe that there are no incoming majority-spin $\Delta_1$ states at the E$_F$ (Figure \ref{Fig:3}). In contrast, for the  minority spin channel in the PC state, we observe tunneling hot-spots around $\Gamma$ point, however showing a much weaker transmission. Apart from the highly conducting channel around the $\Gamma$ point, we observe appearance of conducting channels around the Brillouin zone corners for the majority spin channel showing a four-fold rotational symmetry of the Fe$_2$CrTe electrode. These are identified as resonant tunneling states.\cite{Miura_2008,Butler_2008} In the APC state the spin up and down channels exhibit similar transmission properties with considerably large tunneling states around the $\Gamma$ point (Figure \ref{Fig:11}).

Now as we increase the barrier thickness the transmission around $\Gamma$ point gets highly affected for the majority spin channel, showing astonishingly small transmission around $\Gamma$ and the resonant tunneling states are also diminished. However, transmission for the minority states are not largely affected with increasing barrier thickness (Figure S10).\cite{Supple} This large decrement in the majority spin transmission can be explained form the majority spin band structure of the bulk cubic Fe$_2$CrTe (Figure \ref{Fig:3}), where we do not observe any bands crossing the E$_F$ along the $\Gamma$ - X direction ($i.e$ along the propagation direction (001). These bands are mainly responsible for the highly spin-polarized transmission\cite{Butler_2008, Peter_2009,Miura_2008}, provided that those bands have some preferred orbital character. Due to the absence of such bands along the $\Gamma$-X direction in the majority spin channel, we observe large decrement in the majority spin transmission with increasing barrier thickness (Table \ref{tab:5} and Figure S10\cite{Supple}).

To understand why larger TMR ratio is possible for the tetragonal phase as comapared to the cubic phase, we further investigate the microscopic tunneling process for the tetragonal phase. In Figure \ref{Fig:12}, we have plotted the spin-and k-resolved transmission coefficients at E$_F$, T$^{\sigma}(E_F, K_{||})$ for the tetragonal Fe$_2$CrTe phase in both PC and APC states. This provides a vivid picture of how tunneling process can be relalized in this heterojunction. At a first glance, all the transmission patterns (including parallel and anti-parallel) have a four-fold rotational symmetry, which is in accordance with the C$_{4v}$ symmetry of the MTJ heterostructure. We note that for the majority spin electrons in the PC state, the very sharp transmission features around the $\Gamma$ point domminate the transmission process. However, minority spin electrons in the PC state show $\Gamma$-centric transmission, which are of much weaker intensity. We observe for the tetragonal phase that there is an overall  increase in the transmission of the majority as well as of the minority spins around the $\Gamma$ point in the parallel configuration, as compared to the cubic phase. This is primarily due to the presence of $\Delta_1$ symmetric bands in the bulk tetragonal phase, along the $<001>$ direction. In the APC state, however, we find that there is a dramatic reduction in the transmission of both the spin channels (Figure \ref{Fig:12}), due to a stronger suppression of the electron tunneling proceess, which leads to the larger TMR ratio for the tetragonal phase as compared to the cubic phase.

\section{Conclusion}
In this work, we have carried out a first principles study on the structural, mechanical, electronic and transport properties of a novel full Heusler chalcogenide, Fe$_2$CrTe. The compound is found to undergo a volume conserving tetragonal distortion and a clear minimum has been observed at c/a $\sim$1.26. The cubic phase shows a ferromagnetic behavior with a nearly half metallic character, whereas the tetragonal phase exhibits a ferrimagnetic and fully metallic nature.  This behavior has been found to be robust against a sizable value of Hubbard onsite electron-electron correlation term for both the transition metal atoms. The compound is found to possess no negative phonon frequencies  in both the phases and no martensite phase transition (MPT) was observed, as suggested from the temperature dependent free energy behavior. Further we have also established the mechanical stability of both the phases. We have also studied the effect of uniform strain on the electronic and mechanical properties of the system in cubic phase. It shows 100$\%$ SP on applying a negative uniform strain and the compound is found to be mechanically stable over almost the whole range of lattice constant. To probe the effect of substrate on various physical properties a thin film of 13 mono-layers of Fe$_2$CrTe is placed on a MgO substrate, which shows an energetically stable composite system and the spin polarization has been found to continue to be high for the cubic phase (above 70$\%$). We have further investigated the transmission profile of Fe$_2$CrTe/MgO/Fe$_2$CrTe heterojunction in both the cubic and tetragonal phases. Spin-transport properties for the tetragonal phase looks promising for lower thickness of spacer layer (5 ML). Finally, in light of all the above discussions, synthesis and characterization of the predicted alloy seem essential to understand its structural, electronic, magnetic and spin transport properties and our present work awaits the experimental validation. 

\section{Acknowledgements}
Authors thank the director, RRCAT for facilities and encouragement. We thank A. Banerjee, H. Ghosh and T. Ganguli for scientific discussions. The scientific computing group, computer division of RRCAT, Indore is thanked for the help in installing and support in smooth running of the codes. JB thanks D. Pandey, A. Kumar for useful discussions during the work. JB and RD thank RRCAT and HBNI for financial support.

\bibliography{Reference.bib}

\end{document}